# Study of grain boundary transparency in $(Yb_{1-x}Ca_x)Ba_2Cu_3O$ bi-crystal thin films over a wide temperature, field and field orientation range

*Pei Li*[1,2], *Dmytro Abraimov*[1], *Anatolii Polyanskii*[1], *Fumitake Kametani*[1] and *David Larbalestier*[1].

1. Applied Superconductivity Center, National High Magnetic Field Laboratory, Tallahassee, Florida 32310.
2. Now at: Technical Division, Fermi National Accelerator Laboratory, Batavia, Illinois 60510.

**Abstract**

The residual low angle grain boundary (GB) network is still the most important current-limiting mechanism operating in bi-axially textured rare earth barium copper oxide (REBCO) coated conductors. While Ca-doping is well established to improve supercurrent flow across low angle GBs in weak fields at high temperatures, Ca-doping also depresses $T_c$, making it so far impractical for high temperature applications of REBCO coated conductors. On the other hand, high field magnet applications of REBCO require low temperatures. Here we systematically evaluate the effectiveness of Ca-doping in improving the GB transparency, $r^{GB}= J_c^{GB}/J_c^{grain}$, of low angle $Yb_{1-x}Ca_xBaCuO$ [001] tilt bi-crystal films down to 10K and with magnetic fields perpendicular and parallel to the film surfaces, while varying the Ca and oxygen doping level. Using Low Temperature Scanning Laser Microscopy (LTSLM) and Magneto-Optical Imaging (MOI), we found $r^{GB}$ to strongly depend on the angle between magnetic field and the GB plane and clearly identified regimes in which $J_c^{GB}$ can exceed $J_c^{grain}$ ($r^{GB}>1$) where the GB pinning is optimized by the field being parallel to the GB dislocations. However, even in this favorable situation, we found that $r^{GB}$ became much smaller at lower temperatures. Calculations of the GB Ca segregation profile predict that the high $J_c$ channels between the GB dislocation cores are almost Ca-free. It may be therefore that the positive effects of Ca doping seen by many authors near $T_c$ are partly a consequence of the higher $T_c$ of these Ca-free channels.

## Introduction

While rare-earth-barium-copper-oxide (REBCO) coated conductor is a promising candidate as a practical superconductor for high magnetic field applications, its performance is largely limited by the depressed current-carrying capability of its grain boundaries (GBs). GBs with small misorientation angles greater than ~5° substantially reduce the critical current density ($J_c$) of GBs [1-3], which makes the GB problem a critical one, since real, long-length coated conductors are made on polycrystalline substrates, where low angle GBs are inevitable, despite bi-axial texturing of the substrate [4].

The $J_c$ depression at GB in REBCO thin films was first demonstrated by Dimos *et al.* [1]. $J_c$ of GBs decays exponentially with misorientation angle $\theta$ beyond a critical angle $\theta_c$. Since then, great effort has been devoted to further understand and improve low angle GB properties [2, 3]. Other studies have shown that misorientation angle is not the only variable in the GB problem. The strength and orientation of the magnetic field have a significant effect on GB transparency too. A number of studies [5-7] found that $J_c$ of the GB and the grain converged with increasing field. The field at which $J_c^{GB}$ converged with $J_c^{grain}$ was found to increase with GB misorientation angle $\theta$. Diaz *et al.* [8] found that $J_c^{GB}$ peaks in an angular-dependence study of $J_c^{GB}$, when the magnetic field was aligned with the GB dislocations. The temperature dependence of $J_c^{grain}$ and $J_c^{GB}$ is another important aspect of the GB problem. Studies over a broad temperature range [6] showed that GBs became less transparent at lower temperatures. Among the studies to improve GB transparency, one of the most important findings is the positive effect of Ca-doping, which introduces extra holes that should compensate the reduction of carrier density at the GBs induced by strain and charge inhomogeneity [2, 9, 10]. Detailed microstructural study [7] showed that the underlying reason of improved GB transparency is not simply a uniform Ca doping, because Ca was found to preferentially concentrate at the GBs. Furthermore, the Ca distribution along and across the GB is also highly non-uniform. A segregation model was proposed to explain the observed complex GB Ca distribution [7]. This model shows that Ca segregation is driven by the combined effects of the local strain field [11-13] and the GB charge imbalance [14, 15]. As will be discussed in this paper, a stronger Ca segregation to the GB dislocation is predicted in (YbCa)BCO because Yb is the smallest rare earth element that can form a superconducting RE-123 phase. Since Ca-doping is known to depress $T_c$, a stronger Ca segregation to the GB offers the possibility of improving $r^{GB}$ at a lower overall Ca-doping, which might allow a smaller $T_c$ loss for the doped phase. This was our major motivation for choosing the (YbCa)BCO system in this study.

Because of the interest in high temperature applications of REBCO, the GB properties and the effectiveness of Ca doping have usually been evaluated near $T_c$ in the vicinity of 60-80K. However, with the growing interest in building HTS high-field magnets that operate at low temperature and in magnetic fields of complicated orientation, the property of GBs and the effectiveness of Ca-doping deserve study over a much broader parameter space.

This work presents a systematic study of GB transparency $r^{GB}=J_c^{GB}/J_c^{grain}$ in Ca-doped YbBCO thin films using techniques that can distinguish grain and GB properties. A set of YbBCO thin films with different Ca-doping levels were prepared using Pulsed Laser Deposition (PLD) on [001] tilt $SrTiO_3$ (STO) bi-crystal substrates. In this study, the effect of applied field and its angle to the GB plane, temperature, Ca-doping level and oxygen content established during post-growth anneals were evaluated. Visualization of the GB properties by Low Temperature Scanning Laser Microscopy (LTSLM) [16] and Magneto-Optic Imaging (MOI) [17] was used to study $r^{GB}$. In previous studies, the transparency of the GBs was most usually evaluated by standard four-probe DC transport measurement. In transport methods, $J_c$ is always defined by the "weakest" point in the sample track, even though the actual location of such weak points generally remains unknown. By assuming $J_c^{GB} \leq J_c^{grain}$, previous studies have reported $J_c^{GB} = J_c^{grain}$ with increasing applied field and temperature. On the other hand, there was also speculation that $J_c^{GB}$ could actually be higher than $J_c^{grain}$ due to the strong vortex pinning provided by alignment of applied field with the GB dislocations [8]. However, this was very difficult to verify directly by standard transport measurements. With LTSLM, we were able to unambiguously define the local $J_c^{grain}$ and $J_c^{GB}$ with a spatial resolution of 1-2$\mu$m. This capability allowed us to clearly observe the shift of the local dissipation from the GB to the grain with increasing field and temperature. For the first time, we proved the striking result that $J_c^{GB} > J_c^{grain}$ in samples when measured in perpendicular fields.

Nevertheless, Ca-doping was found to be less effective at improving $r^{GB}$ under conditions that are most relevant to high field applications, *i.e.* low temperatures and fields with non-perpendicular components. This observation is discussed in the context of the strong pinning provided by GB dislocations and the predictions of the GB segregation model. Together these considerations provide further evidence that planar grain boundaries have special characteristics that must be carefully considered when seeking general understanding of HTS GBs over a wide parameter space.

**Sample preparation and experimental details**

**Film deposition**

The thin $(Yb_{1-x}Ca_x)BaCuO$ films were grown using a LPX-200 248nm excimer pulsed laser deposition (PLD) system with targets of $x$=0, 0.1 and 0.3. Films were grown on 6º and 7º [001] tilt $SrTiO_3$ (STO) bi-crystals. Most of the purchased bi-crystal substrate GBs contained random point-like defects (with a typical size of several $\mu$m) which resulted in non-uniform GBs in the superconductor films grown. As described below, such defects do not pose a problem for LTSLM study because the narrow sample bridge can always be placed in defect-free regions. However, for the magneto-optical imaging (MOI) study, such defects strongly distort the MO discontinuity line pattern, making the extraction of $J_c^{GB}/J_c^{grain}$ ratio very complicated. We were able to grow a high quality $x$=0.3 film using a defect-free 6º bi-crystal substrate, which was used for intensive MOI study. Using this sample, we were able to repetitively measure $r^{GB}$ of the same sample after a series of anneals with different oxygen partial pressures, without any problem of contact pads degrading the induced current pattern. In addition to bi-crystal samples, some films were also grown on STO single crystal substrates. The bi-crystal films were used to study GB properties (using LTSLM and MOI) and the single crystal films were mainly used as benchmarks to understand the effect of oxygen annealing and Ca-doping level on $T_c$ and the intra-grain $J_c$. The optimized film deposition parameters are listed in Table 1.

The substrates were glued on an inductive heater block using silver paste and loaded into the PLD chamber. The chamber was then pumped to a vacuum of $10^{-6}$mtorr before deposition. After deposition, the PLD chamber was backfilled with oxygen to 100torr as the heater temperature was reduced to 500ºC. At 500ºC, the chamber was then filled with oxygen to 1 bar and the film was *in situ* annealed for 1 hour. *Ex situ* annealing was done in either 1 bar flowing oxygen or under a reduced oxygen pressure, as described later. The detailed information on the major sample set discussed in this paper is summarized in Table 2.

The resulting films had thicknesses of 200nm-400nm. The thickness of each tested film was measured by a Focused Ion Beam (FIB) cross-section cut. Sample tracks were patterned with FIB and gold contact pads were placed using DC sputtering.

**Patterning of single GB tracks**

Single GB tracks for $r^{GB}$ evaluation were patterned using a FIB. To ensure reliable results, a single GB track should not contain any local inhomogeneity other than a well-defined GB. The positions of the single GB tracks were carefully chosen to avoid major intra-grain particulates, which are common in PLD films. An example of such a carefully placed single GB track is shown in Figure 1.

**Local transport $J_c$ measurement using LTSLM**

The low temperature scanning laser microscope (LTSLM) is a hot spot probe technique [16, 18-20] which utilizes a focused probe laser beam to scan the sample surface to induce local temperature perturbations while the voltage change $\delta V$ responses across the sample are recorded as a function of scan position $(x,y)$. The spatially resolved $\delta V(x,y)$ responses reflect the local property variation of the superconductor samples. Detailed descriptions of the principles of LTSLM and our setup can be found in [16, 21]. Here we focus on two applications of LTSLM that are relevant to this study, the measurement of local $J_c$ at $T<T_c$ and visualization of the local grain orientation by measurement of thermoelectric power at room temperature.

The principle of local $J_c$ measurement in a 1D superconductor track is illustrated in Figure 2. In a 1D track with a constant cross section $A$, the $I_c$ and $J_c$ are assumed to vary only along the current flow direction, i.e. $J_c=J_c(x)=I_c(x)/A$. A schematic $I_c(x)$ distribution is plotted in Figure 2. A bias current $I_{bias}$ is fed through the superconductor link which is kept at $T<T_c$. When $I_{bias}$ is below the entire $I_c(x)$ curve ($I_{bias1}$ in Figure 2), the laser perturbation does not induce any $\delta V$ response. As $I_{bias}$ reaches the lowest point ($I_{bias2}$ in Figure 2) of the $I_c(x)$ curve, the first $\delta V$ response to perturbation by the laser appears. Further increasing $I_{bias}$ causes more parts of the link to generate $\delta V$ responses. It can be also seen from Figure 2 that segments with lower $J_c$ will generate stronger $\delta V$ responses at a constant $I_{bias}$. In our interpretation of the LTSLM results, the local $I_c(x)$ is defined by the lowest $I_{bias}$ current that generates a detectable local $\delta V$ response. It should be noted that comparison of $I_c$ thus defined by LTSLM and that by standard transport methods at $1\mu V/cm$ shows that $I_{bias}\sim110\%I_c(x)^{transport}$ is needed to generate detectable $\delta V(x)$ responses above the noise level.

The detection of grain-to-grain misorientation with LTSLM is based on the fact that the high temperature superconducting cuprates have layered structures with strongly anisotropic Seebeck coefficients [20]. The part of a cuprate sample perturbed by the laser beam acts like a thermopile and the induced thermoelectric voltage $\delta V$ response is a convenient indicator of grain orientation. An example is given in Figure 5(B), where the two single grains forming the bi-crystal show different contrast due to their misorientation. This measurement does not require bias current and is usually done at room temperature to conveniently locate positions of GBs and other inhomogeneities [21].

As shown in Figure 3, our setup enables the $\delta V(x,y)$ response distribution to be visualized with field applied either perpendicular or parallel to the film surface. Our LTSLM has an applied field capability of 0-5 Tesla using a superconducting magnet.

**Other characterization methods**

A SQUID and a PPMS were used to measure the $T_c$ and $J_c$ of some single crystal samples. SEM and EDS were used to monitor the morphology and elemental compositions of films.

**Results**

**Superconductivity of (Yb$_{1-x}$Ca$_x$)BCO thin films and the effect of oxygen annealing**

Single crystal films with $x$=0, 0.1 and 0.3 were annealed in 1 bar flowing oxygen at 475°C. $T_c$ measurements between anneals were done in a SQUID. The results are shown in Figure 4. With long enough annealing time, $T_c$ of all three films saturated. The stable $T_c$ values were 83.3K, 80.5K and 75.8K for the $x$=0, 0.1 and 0.3 films, respectively. The Ca-doped YbBCO films behaved quite differently compared to the pure ($x$=0) film. First of all, with extending annealing time, the $T_c$ of the Ca-doped YbBCO films increased significantly, while the $T_c$ of the pure film decreased by 3.5K. Secondly, the Ca-doped films needed a significantly longer time to reach a stable $T_c$. The $T_c$ of the pure film stabilized after 5 hours while that of the $x$=0.1 and 0.3 films stabilized only after 15 and 25 hours respectively. The transport $T_c$ and $J_c$ of an $x$=0.1 single crystal film was measured between anneals. The $J_c$ (0.85$T_c$, self-field) of the film increased from 0.167MA/cm$^2$ to 2.68MA/cm$^2$ after 42 hours annealing. No further $T_c$ or $J_c$ changes were found with further annealing.

After reaching stable $T_c$ values after 1 bar oxygen annealing, the Ca-doped YbBCO films were annealed in reduced oxygen pressures. The anneal at each pressure was repeated several times with intermediate SQUID measurements to ensure that $T_c$ had stabilized. With reducing oxygen pressure, $T_c$ of both $x$=0.1 and 0.3 films first increased and then decreased, showing a parabolic dependence on oxygen pressure as the under-doped state was approached. The observed peak transition temperature $T_c^{max}$ was 84.3K for $x$=0.1 films and 79.2K for $x$=0.3 films. The stable $T_c$ at different oxygen pressures were fitted using the universal equation for cuprates $T_c/T_c^{max}$=1-82.6($p$-0.167)$^2$ [23], where $p$ is the number of holes per CuO$_2$ plane, as shown in Figure 4(c).

The morphology and elemental compositions of these samples were monitored with SEM and EDS after each annealing step. Annealing in either 1 bar or reduced oxygen pressure was not found to cause any observable change in morphology or cation composition.

**Study of the GB transparency r$^{GB}$ in bi-crystal samples**

**In field LTSLM results on 7° [001] tilt bi-crystal GBs with various Ca-doping levels and oxygen annealing history**

The r$^{GB}$ of five 7° bi-crystal films with different Ca-doping levels and annealing histories was evaluated using LTSLM. T-0.1 was a 10% Ca-doped film which was *ex situ* annealed in 1 bar oxygen for 2 hours. S-0, S-0.1 and S-0.3 (with 0%, 10% and 30% Ca doping) were films annealed extensively in 1 bar oxygen until their $T_c$ saturated. R-0.3 was a 30% Ca-doped film which was first annealed extensively in 1 bar oxygen and then deoxygenated in 2torr oxygen. The detailed information on these samples is summarized in Table 2.

The LTSLM images from the sample T-0.1 (10% Ca-doped, annealed in 1 bar oxygen for 2 hours) are shown in Figure 5. All the $\delta V$ response images are aligned. The position of the GB can be clearly identified in the room temperature thermoelectric image as the two grains connected by the GB show different contrasts. At 0.85$T_c$ (which corresponds to 77K for YBCO with 90K $T_c$) with a bias current of 33mA, it is very striking that $\delta V$ responses have appeared in most parts of the track but not at the GB, indicating that $J_c^{GB}$>$J_c^{grain}$, *i.e.* r$^{GB}$ is greater than unity. In experiments, intra-grain responses at 0.85$T_c$ first appeared at $I_{bias}$=19.5mA. As a result, r$^{GB}$(0.85$T_c$, self-field) > 1.7. On the other hand, when the temperature was decreased to 0.7$T_c$ (self-field), the GB became the weakest part of the track, as indicated by the strong localized GB response. r$^{GB}$(0.7$T_c$, SF)=0.65.

The LTSLM images from sample S-0.1 (10% Ca-doped, annealed in 1 bar oxygen extensively) are shown in Figure 6. All the images were taken at $0.85T_c$ in 0-4T fields applied perpendicular to the film surface. In self-field, GB was the weakest point with $r^{GB}$=0.65. With increasing field, $r^{GB}$ increased to unity around 2-3T.

The $r^{GB}$ of films that were extensively annealed in 1 bar oxygen, *i.e.* S-0, S-0.1 and S-0.3 (0%, 10% and 30% Ca-doped, annealed extensively in 1 bar oxygen), are plotted as function of reduced temperature $t=T/T_c$ and applied field (perpendicular) in Figure 7. Figure 7 shows several consistent features. For different Ca-doping levels, $r^{GB}$ always increases with increasing applied field and higher reduced temperature $t$. Ca-doping improves $r^{GB}$ and 10% Ca doping generally leads to the highest $r^{GB}$ among the three tested doping levels. In another set of experiments, we imaged the 30% Ca-doped sample S-0.3 in both perpendicular field and parallel field. The $r^{GB}$ results are plotted in Figure 8. Comparing these two field orientations, it is very clear that the GB is less transparent in parallel field. Another major difference is that while $r^{GB}$ increases with increasing perpendicular field, it decreases monotonically with increasing parallel field up to the maximum field of the LTSLM (5T).

The LTSLM results from the oxygen under-doped sample R-0.3 (30% Ca-doped, annealed in 2 torr oxygen) are shown in Figure 9. The low temperature $\delta V$ image shown was taken at 4T (perpendicular field) and $0.7T_c$. The absence of a $\delta V$ response at the GB indicates a transparent GB for which $J_c^{GB}>J_c^{grain}$. Actually, $r^{GB}(0.7T_c, 4T)=1.47$. All the tested $r^{GB}$ data points are plotted in Figure 9. Comparing sample R-0.3 (Figure 8) and sample S-0.3 (Figure 7), the $r^{GB}$ of the oxygen under-doped R-0.3 sample is higher than that of the fully-oxygenated S-0.3 in perpendicular field. $r^{GB}$ of sample R-0.3 also increases faster with increasing perpendicular field. On the other hand, a drastic $r^{GB}$ decrease with increasing parallel field is observed in R-0.3. In parallel field, the discrepancy between $J_c^{GB}$ and $J_c^{grain}$ also grew faster in R-0.3 with decreasing reduced temperature $t$. At temperatures lower than $0.85T_c$, the average electric field across the track reached 1000$\mu$V/cm before any intra-grain $\delta V$ response could be detected. As a result, for sample R-0.3, we were only able to obtain $r^{GB}$ data at $0.85T_c$ in parallel field.

**MOI study of a 6º GB (Yb$_{0.7}$Ca$_{0.3}$)BCO bi-crystal film.**

An $x$=0.3 film grown on a high quality, defect-free 6º STO bi-crystal substrate was imaged at 10K using MOI. Magneto-Optical patterns with well-developed bright discontinuity lines were taken at different stages of oxygen annealing at 475ºC and some of the field-cooled (FC) images are shown in Figure 10. The $J_c^{grain}$ and $J_c^{GB}$ were calculated by quantitative analysis of the MOI images using the method proposed in Ref [17], where the ratio of $r^{GB}= J_c^{GB}/J_c^{grain}$ was determined from the angle made by the discontinuity lines with the GB and the long side of the bi-crystal film. When the sample was annealed for only 2 hours in 1 bar oxygen (Figure 10(a)), a well-developed, single "roof" pattern was observed. This shows that the GB between the two grains is not obstructing current flow. MO contrast along the GB is absent, which is evidence of no preferred flux penetration along the GB. The pattern shows a strong coupling across the GB and a high $r^{GB}$ close to unity. With a longer annealing of 36 hours in 1 bar oxygen (Figure 10(b)), the MO image evolved into two decoupled "roof" patterns. The GB is then an obstacle to current flow and an easy path for magnetic flux penetration. These suggest that long oxygen annealing increases the difference between $J_c^{GB}$ and $J_c^{grain}$. Though both $J_c^{GB}$ and $J_c^{grain}$ increase, $r^{GB}$ drops to 0.46. More interestingly, an additional annealing in a reduced oxygen pressure (500mtorr) partially recovered the GB transparency for current flow and $r^{GB}$ increased to 0.68 (Figure 10(c)). The tiny domain in the bottom grain in Figure 10(b) and 10(c) was due to contamination introduced by sample handling. This does not affect the above interpretation of the MOI results as the contamination was not immediately close to the GB and SEM observation showed that the affected region remained confined in subsequent heat treatments.

**Discussion**

This study of $r^{GB}$ has clearly shown that, along with GB misorientation angle, other variables also strongly affect $r^{GB}$. These variables include the strength and the orientation of applied field, the measurement temperature, the Ca-doping level and the intra-grain critical current density $J_c^{grain}$.

**Effect of post-growth annealing on the intra-grain properties and $r^{GB}$**

While this study is mainly about GB properties, attention has to be paid to intra-grain properties too for the following reasons: (i) the evaluation of GB transparency $r^{GB}$ involves both intra-grain and inter-grain properties, which is especially true for Ca-doped REBCO as the beneficial effect of Ca-doping is attributed to an enhanced carrier density provided by the introduction of extra holes; (ii) from the point of view of application, both high $J_c^{grain}$ and high $r^{GB}$ are desired.

An unusual finding of our study is the long time needed to establish a stable $T_c$ and $J_c^{grain}$ in the as-grown Ca-doped films during annealing in 1 bar oxygen at 475°C. The effect of post-growth annealing of PLD REBCO films in 1 bar oxygen is generally considered to be further oxidation through oxygen diffusion. Previous studies show that oxygen becomes mobile in REBCO above 250-300°C. The oxygen diffusion coefficients in YBCO are highly anisotropic [24]. In single crystal YBCO, the in-plane diffusion coefficient $D_{ab}$ is of order $10^{-12}$ cm$^2$/s in the temperature range of 400-500°C, which is at least three orders of magnitude greater than that along the *c*-axis, $D_c$ (~$10^{-15}$ cm$^2$/s) [24, 25]. In PLD-films, the effective diffusion coefficient was found to be less than an order of magnitude smaller than $D_{ab}$, but much greater than $D_c$ of a single crystal [26], suggesting that despite a predominant c-axis texture, the island growth leaves enough diffusion channels along the c-axis such as island junctions and/or threading dislocations, all of which contribute to enhance $D_c$. $10^{-12}$cm$^2$/s and $10^{-13}$cm$^2$/s can thus be used as the upper and lower limits to estimate the effective oxygen diffusion depth *d* in our (Yb$_{1-x}$Ca$_x$)BCO films. An hour of annealing at 475°C corresponds to *d* between 268nm and 840nm, comparable to the typical thickness of 200-400nm of our films. In most reported work on PLD REBCO films [27, 28], several (less than 5) hours were found to be enough for full oxidation to a stable $T_c$ and $J_c$, which is consistent with our observation with pure YbBCO films.

The trend of $T_c$ variation of Ca-doped films during the post-growth annealing in 1 bar oxygen is different from that of pure films. As demonstrated by many studies, pure PLD REBCO films grown using similar procedures were found not to be strongly under-doped. In this sense, if the only effect of annealing in 1 bar oxygen is further oxygen-loading, $T_c$ is expected to decrease due to over-doping. However, in this study, this was only found in pure YbBCO films. On the contrary, the $T_c$ of *x*=0.3 films increased monotonically from 43K to 75.8K. The $T_c$ of the *x*=0.1 film first decreased from 71K to 65K and then increased slowly to 80.5K. Most reported post-growth annealing of Ca-doped REBCO films is limited to several hours and a considerable scatter of $T_c$ values is found in the literature [6, 29, 30]. Several works [31, 32] that extended annealing time did report similar $T_c$ increases. While the cause of the unexpected $T_c$ and $J_c^{grain}$ variation of the Ca-doped films during post-growth annealing requires further investigation, our study has clearly shown that Ca-doped YbBCO required extensive annealing in 1 bar oxygen to establish a stable $T_c$ and to optimize $J_c^{grain}$. However, after the Ca-doped films did stabilize in 1 bar oxygen, their $T_c$ and $J_c^{grain}$ changes caused by annealing in reduced oxygen pressure were consistent with the reported behavior of over-doped REBCO, as is shown in Figure 4(c).

Our LTSLM and MOI results showed that $r^{GB}$ of Ca-doped GBs did not increase with $J_c^{grain}$. On the contrary, $r^{GB}$ was higher when $J_c^{grain}$ was not optimized. This indicates that $J_c^{GB}$ is affected differently by oxygen annealing compared with $J_c^{grain}$. It should be pointed out that the correlation between high $r^{GB}$ and reduced $J_c^{grain}$ is not unique to Ca-doped samples. The condition $r^{GB}>1$ was also observed by LTSLM in a pure ($x=0$) bi-crystal film after being annealed in reduced oxygen pressure when measured in perpendicular fields.

**Effect of the applied field orientation with respect to the GB dislocation axes**

The drastically different $r^{GB}$ behavior in perpendicular and in parallel fields (Figure 8 and Figure 9) well supports the idea that the GB dislocation cores play an important role in GB flux pinning [8]. In PLD REBCO films grown on *c*-axis textured substrates, due to the nature of island growth mode, the GB planes and the dislocation cores are predominantly perpendicular to the film surface. In perpendicular field, the flux lines are aligned with the GB dislocations, making them effective pinning centers. For Abrikosov-Josephson vortices pinned by GB dislocations, the pinning force density, $F_{pinning}=J_c \times H$, scales with applied field $H$ as $H^{1/2}$ [33, 34]. The pinning force densities at the 7° GBs, $F_{pinning}^{GB}$ in S-0, S-0.1 and S-0.3 are calculated and plotted on double-logarithmic scale in Figure 11. For fields between 0.5T and 5T, and temperatures $t$ between $0.3T_c$ and $0.7T_c$, each set of data points form straight lines. The slopes of these lines fall between 0.4-0.65, thus the $F_{pinning}^{GB} \propto H^{0.4-0.65}$, which is consistent with the scaling behavior of pinning by dislocations. Our LTSLM visualization study has further shown that, with a reduced intra-grain pinning, the pinning from the dislocation cores was strong enough to cause $J_c^{GB}>J_c^{grain}$ in perpendicular field when $J_c^{grain}$ was less than fully optimized. This is also corroborated by the MOI observation of the 30% Ca-doped 6° GB (Figure 10), where $r^{GB}$ is higher when $J_c^{grain}$ is not optimized.

When the applied field deviates from being parallel to the GB dislocations, the pinning by the GB dislocations loses its effectiveness. As shown in Figure 8 and 9, $r^{GB}$ decreases monotonically with increasing field in parallel field. This is because the GB dislocations cannot provide any position-dependent pinning potential when the flux line is parallel to the *ab* plane. Furthermore, due to the layered structure of cuprates, flux lines preferentially lay between the $CuO_2$ planes, where the order parameter is low, and flux motion across the planes needs to overcome high energy barriers. This is known as intrinsic pinning [35]. As a result, $r^{GB}$ is further reduced. It is also found here in parallel field that a depressed intra-grain pinning in oxygen under-doped samples causes $r^{GB}$ to decrease even faster with increasing field than in fully-oxygenated samples.

**Ca segregation and temperature dependence of $r^{GB}$**

As described in the introduction, one of the major motivations to study $(Yb_{1-x}Ca_x)BCO$ was to explore the effects of a stronger Ca segregation to GBs predicted by the strain-charge model [7]. This model successfully described the observed non-uniform Ca distribution along and across GBs in Ca-doped YBCO. In REBCO with radius of $RE^{3+}$ smaller than that of $Ca^{2+}$, under the effect of GB strain field $p(x, y)$ [11], the concentration of Ca dopant is energetically favored in the tensile parts of the dislocation cores. The GB electric potential $\phi(x, y)$ further modifies the Ca distribution across GBs. The combined effect of strain and electric field results in a Ca segregation profile both across (*x* direction in Figure 12) and along (*y* direction in Figure 12) GBs, as shown in Figure 12. Mathematical details of the model can be found in Ref [7]. Comparing with $Y^{3+}$, the larger volume difference between $Yb^{3+}$ and $Ca^{2+}$ is expected to achieve similar $r^{GB}$ improvement at a lower overall Ca-doping level. Using the same fitting parameters as Ref [7] and considering the different size of $Yb^{3+}$ and $Y^{3+}$[36], Ca distributions along a 7° GB in $(Yb_{0.7}Ca_{0.3})BCO$ and $(Y_{0.7}Ca_{0.3})BCO$ are plotted in Figure 12. In $(Yb_{0.7}Ca_{0.3})BCO$, the peak Ca concentration level along the GB plane (x=0 in Figure 12) is more than 2 times the bulk level, about one third higher than that of $(Y_{0.7}Ca_{0.3})BCO$. Our results clearly confirm a GB transparency improvement in the high temperature regime. As shown in Figure 7, in most of the parameter space covered by this study, 10%Ca doping has the best performance and 30%Ca-doping is not better than 10%Ca-doping. On the other hand, 30% Ca doping was found to be optimal for GBs in YBCO [11]. These facts imply the existence of an optimum Ca doping level for YbBCO which is lower than that for YBCO. Though the reason why excessive Ca reduces $r^{GB}$ is unclear, it might be caused by Ca substitution at sites other than RE sites. Earlier study of Ca-doped YBCO samples fabricated by solid-state-reaction reported that the maximum Ca substitution at Y sites was no more than 16% [21, 37].

In the low temperature regime, though the benefit of Ca doping could still be observed, it became less pronounced. The computed GB Ca segregation could also explain the fast $r^{GB}$ drop with decreasing temperature. As shown in Figure 12, a major feature of the predicted Ca distribution is the existence of supercurrent channels with markedly reduced Ca concentration. Parts of these channels have zero Ca content. As a result, these channels have $T_c$ higher than that of the grain on either side. The $J_c$ of REBCO has a temperature dependence of $(1 - t)^\alpha$ [38]. The index $\alpha$ is temperature-dependent. At high temperature, $\alpha = 1$, while at low temperature $\alpha$ varies between 1.5 and 3, depending on substrate material and film fabrication techniques. A higher GB $T_c$ value causes the Ca-deficient current channels to have higher $J_c$ than the grain when measured at the same *absolute* temperature $T$. As $J_c$ scales with reduced temperature $t$, these supercurrent channels have a higher $T_c' = T_c+\delta T_c$, which is equivalent to a lower reduced temperature $t' = T/(T_c + \delta T_c)$. When evaluated at the same *absolute* temperature $T$, the difference in reduced temperature for grain and GB is then $t – t' = T \cdot \delta T_c/(T_c+\delta T_c)$, which is proportional to $T$. As a result, the benefit from Ca segregation is most pronounced at higher temperature and becomes less important at lower temperature. For example, using the stable $T_c$ values shown in Figure 4(a), the channels in a 10% Ca-doped film are found to carry $J_c^{channel}/J_c^{grain}$=120% at $0.85T_c$, assuming $\alpha$=1. It is plausible that this $J_c$ enhancement is due to local $T_c$ imbalance caused by Ca segregation. On the other hand, the ratio $J_c^{channel}/J_c^{grain}$ decreases sharply for $t <0.7$ and becomes negligible at lower temperatures. Another reported result of Ca segregation is the expansion of non-superconducting dislocation core regions due to the local Ca content maxima [7]. When a Ca-doped GB is measured at high temperature, the shrinking of current channels is compensated by gain in $J_c^{GB}$ caused by $T_c$ difference. Below $t$=0.5, the gain from the higher supercurrent channel $T_c$ decreases significantly and the negative effect of dislocation core expansion becomes more important. Detailed TEM study of these $(Yb_{1-x}Ca_x)BCO$ GB samples is actively in progress and initial results have confirmed the Ca segregation to the GBs and the existence of low Ca channels [39].

**Conclusion**

In this paper, we studied some major factors that affect the transparency of low angle GBs in Ca-doped YbBCO films. From our results, several lessons could be learned. First of all, GB transparency depends on both $J_c^{grain}$ and $J_c^{GB}$, which change in different ways during post-growth annealing. Furthermore, $J_c^{GB}$ can actually be higher than $J_c^{grain}$, which is attributed to a combined effect of pinning by GB dislocation and depressed $J_c^{grain}$ for the orientation $H//c$. In this sense, techniques that can independently and unambiguously define the two quantities are important for the study of GBs. Secondly, pinning by dislocation cores makes a significant contribution to $J_c^{GB}$ in perpendicular field.

On the other hand, while perpendicular field configuration is preferential for r$^{GB}$, other field orientations should also be included in a comprehensive GB study because it does not appear that these orientations show such favorable effects as the *H*//c orientation. In summary, we find that the effectiveness of Ca-doping is limited to temperatures close to $T_c$ and for perpendicular fields. For high field application at low temperature, Ca-doping is not a very useful solution to the GB problem.

**Acknowledgement**

This work is performed at the National High Magnetic Field Laboratory, which is supported by National Science Foundation Cooperative Agreement No. DMR-1157490. We are also grateful to Dr. Alexander Gurevich of Old Dominion University for very useful discussions of his strain and charge segregation model in the initial stages of planning this experiment.

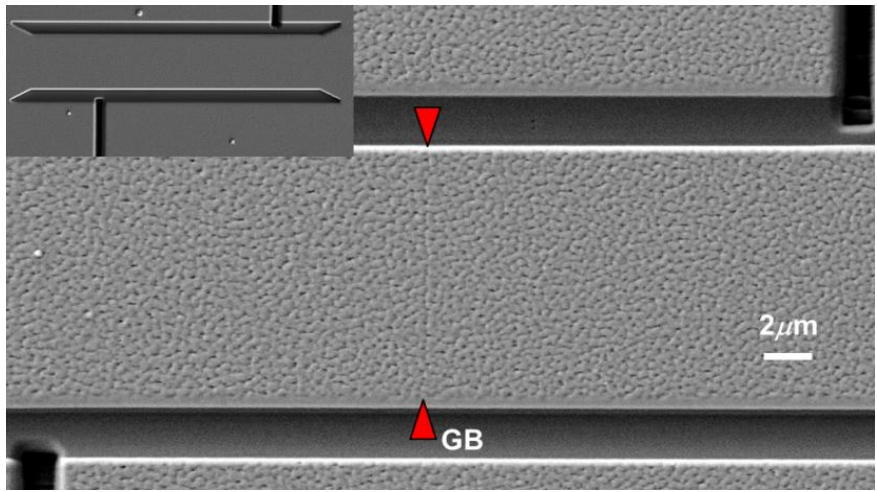

Figure 1. A single GB track patterned with a FIB. The 7º GB is marked by red arrows. The inset shows the entire track. The track does not contain major particulates and the GB is the only inhomogeneity in the track.

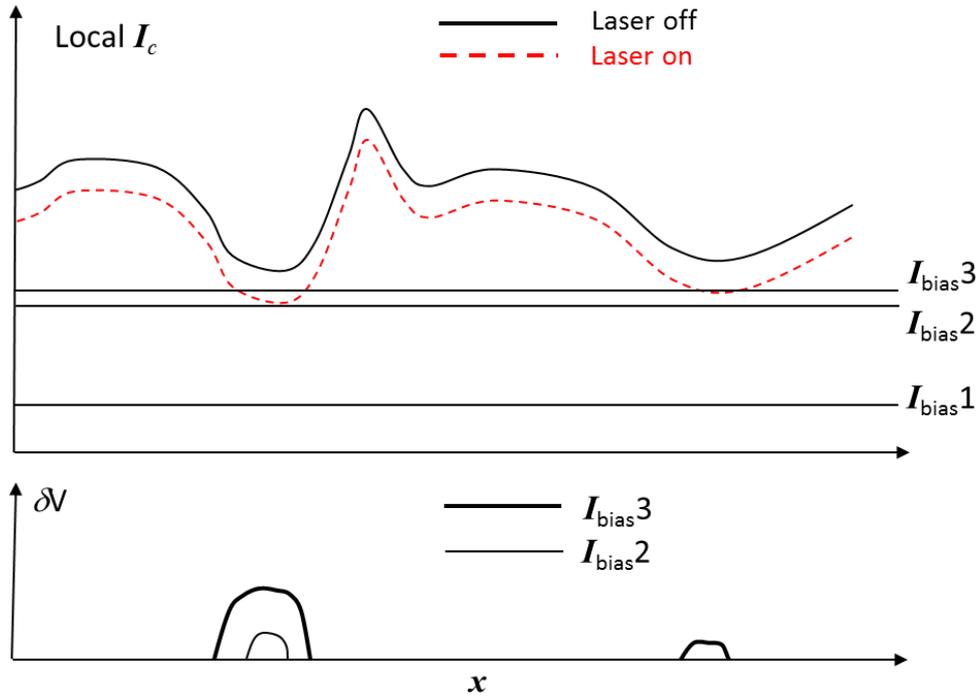

Figure 2. Basic principle of local $I_c$ measurement in a 1D superconductor track using LTSLM [19]. Top: schematic local $I_c(x)$ distribution in a 1D track with and without laser perturbation. The small perturbation from the scanning laser beam slightly shifts the $I_c(x)$ curve. Bottom: $\delta V$ response distributions at different bias currents. $\delta V(x)>0$, if $I_{bias}>I_c(x)$.

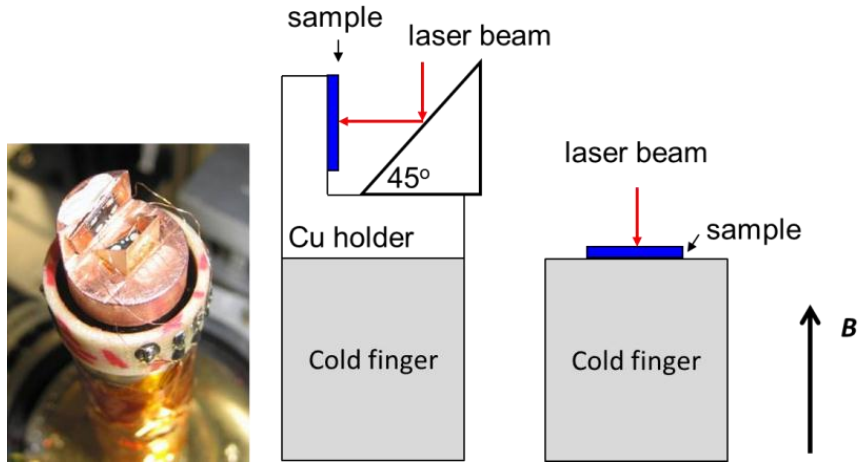

Figure 3. The two visualization modes used in the LTSLM. Left: using a 45° mirror, $\delta V$ distribution can be visualized with the applied field parallel to the film sample surface. Right: $\delta V$ distribution is visualized with the applied field perpendicular to the film surface.

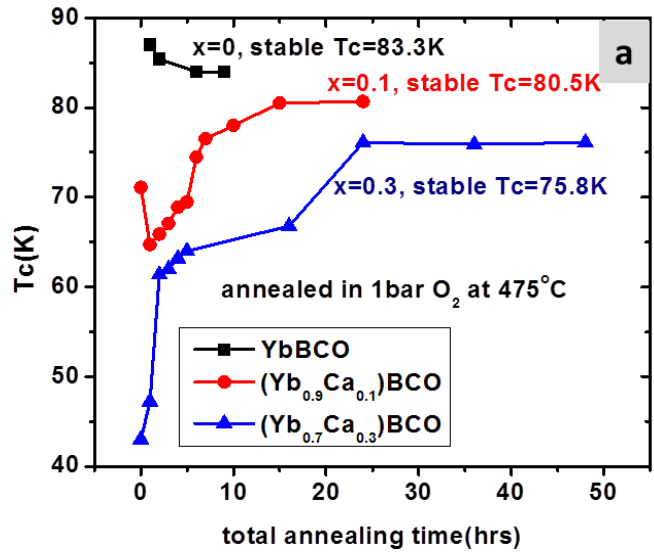

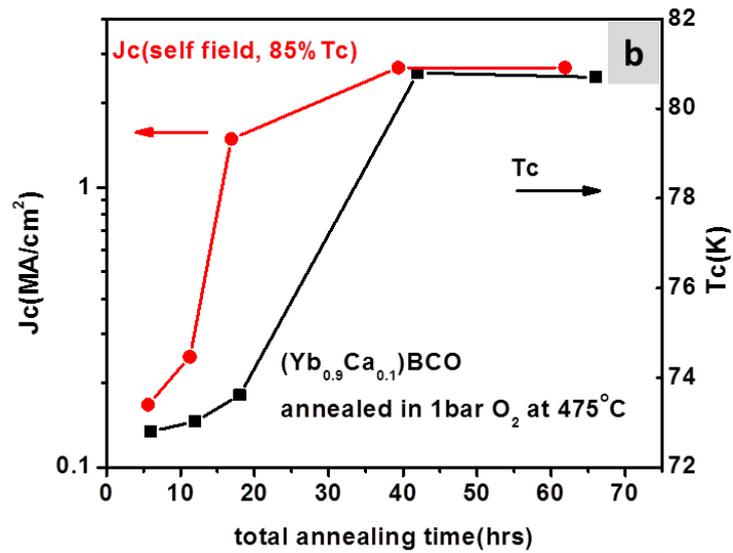

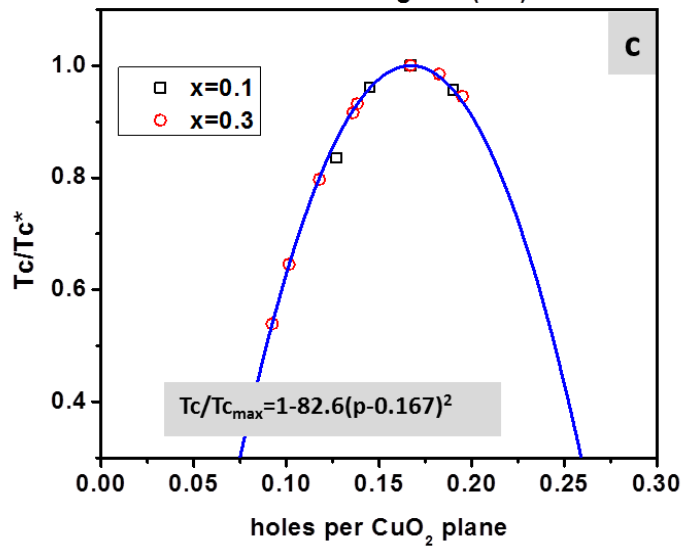

Figure 4. The effect of annealing time and oxygen pressure on (Yb$_{1-x}$Ca$_x$)BCO single crystal thin films. Annealing in 1 bar oxygen at 475°C: (a) $T_c$ dependence on total accumulated annealing time. $T_c$ was measured in a SQUID. (b) Transport $T_c$ and $J_c$ of an $x=0.1$ single crystal film with different annealing times. (c) $T_c$ variation with annealing oxygen pressure. Stabilized $T_c$ values are normalized to their maximum transition temperature $T_c^{max}$ and fitted to equation $T_c/T_c^{max}=1-82.6(p-0.167)^2$ [21].

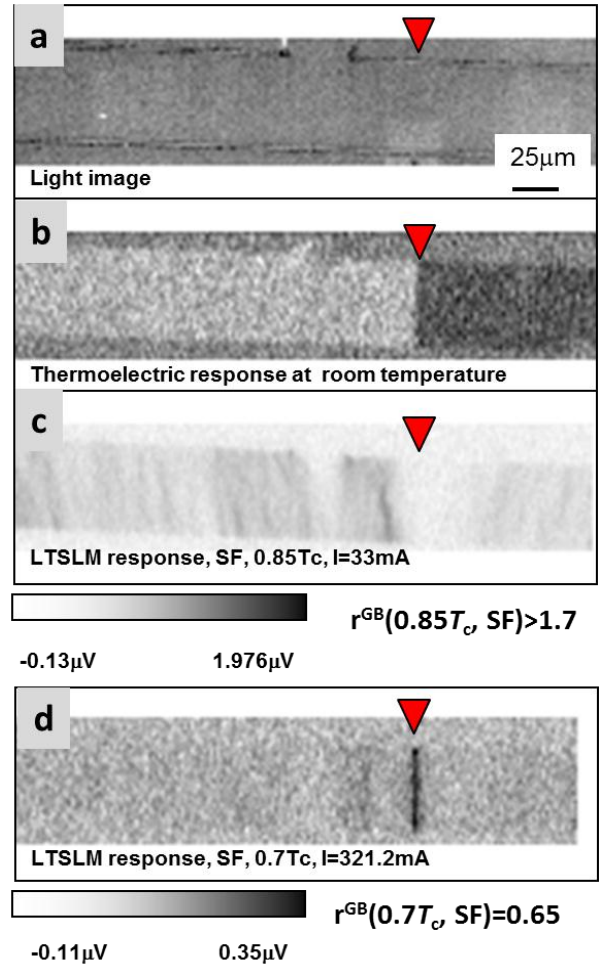

Figure 5. Aligned LTSLM images of an $x$=0.1, 7° bi-crystal film sample T-0.1. This film was annealed in 1 bar oxygen at 475°C for 2 hours after deposition. The $T_c$ and intra-grain $J_c$ (self-field, 0.85$T_c$) of this sample was 61.3K and 0.14MA/cm$^2$ respectively. (a) Light image showing the sample track; (b) room temperature thermoelectric response image showing the position of the 7° GB; (c) $\delta V$ response at 0.85$T_c$, self-field (SF) and $I_{bias}$=33mA ( $J$=0.24MA/cm$^2$). $\delta V$ response appeared at most part of the track but not at the GB, indicating that $J_c^{GB}$>$J_c^{grain}$. (d) $\delta V$ response image at 0.7$T_c$, self-field and $I_{bias}$=321.2mA. The strong localized GB response indicates that $J_c^{GB}$<$J_c^{grain}$. The position of the 7° GB is marked with the red arrows.

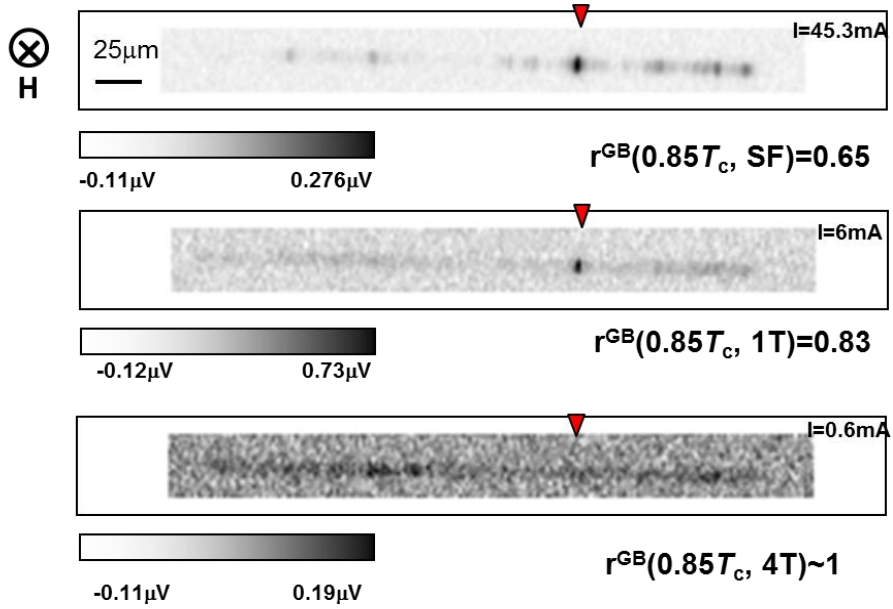

Figure 6. Aligned LTSLM images of an $x$=0.1, 7° bi-crystal film sample S-0.1 at $0.85T_c$, in different perpendicular fields: (a) self-field, (b) 1T, (c) 4T. This film was annealed in 1bar oxygen at 475°C for 24 hours after deposition. The $T_c$ and $J_c^{grain}$ (self-field, $0.85T_c$) of the sample was 79.2K and 1.8MA/cm$^2$ respectively. With increasing field, $r^{GB}=J_c^{GB}/J_c^{grain}$ increased from 0.65 to 1. Position of the GB is marked with red arrows.

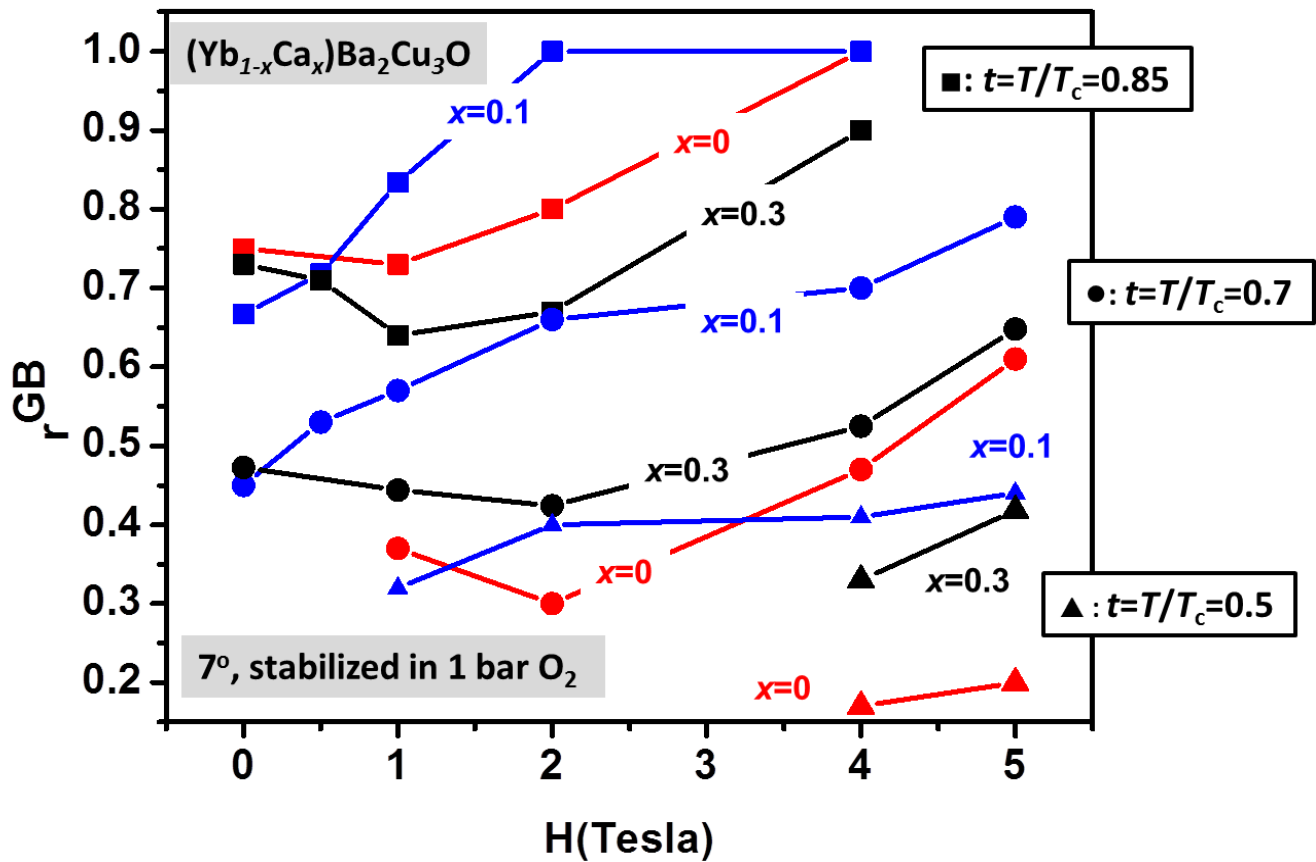

Figure 7. Dependence of GB transparency $r^{GB}=J_c^{GB}/J_c^{grain}$ on Ca-doping level *x*, applied field (perpendicular) and reduced temperature $t=T/T_c=0.85$, 0.7 and 0.5. All the plotted data points are from 7° bi-crystal GBs. All the films were extensively annealed in 1bar oxygen. $r^{GB}$ increases with increasing applied field (perpendicular) and increasing reduced temperature *t*.

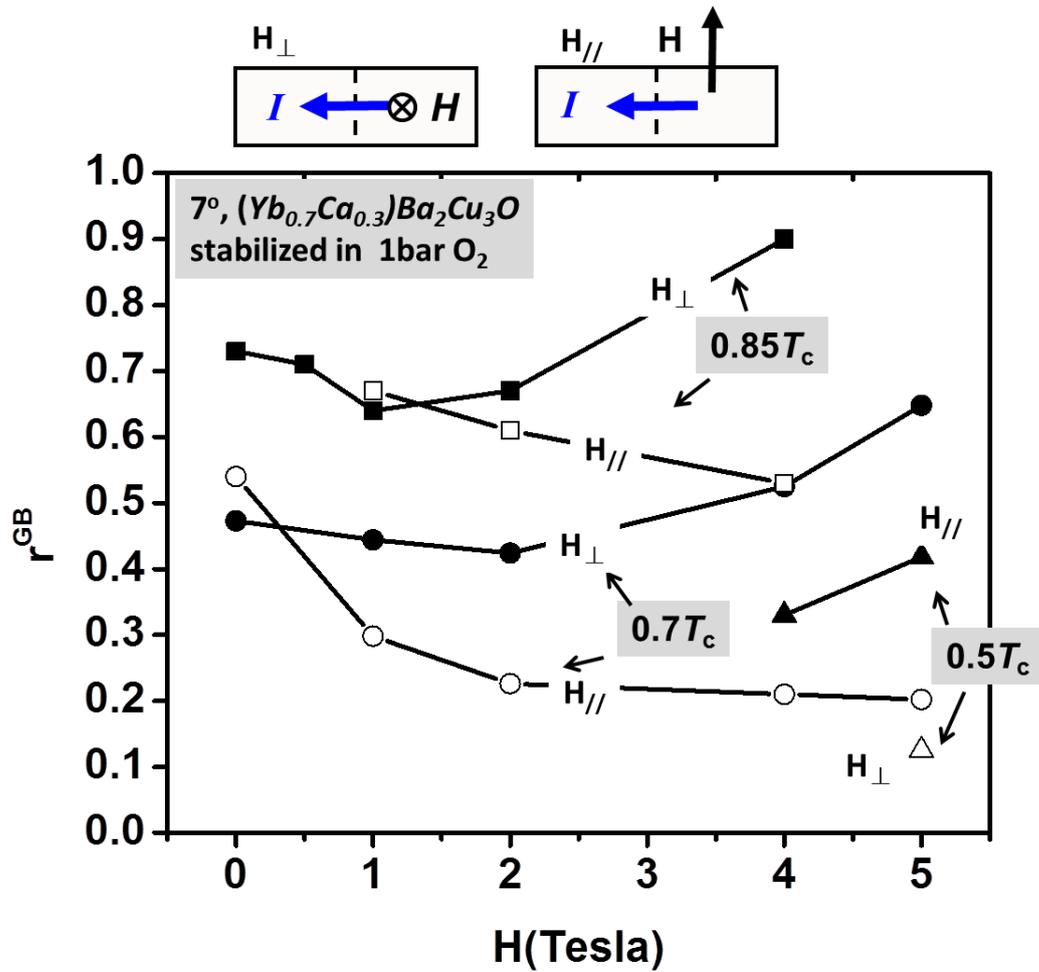

Figure 8. Dependence of the GB transparency $r^{GB}=J_c^{GB}/J_c^{grain}$ on field orientation. The sample was a 7°, $(Yb_{0.7}Ca_{0.3})$BCO bi-crystal film sample, S-0.3. The $T_c$ and $J_c$ of the sample were stabilized by extensive annealing in 1 bar oxygen at 475°C. In perpendicular field, $r^{GB}$ increases with applied field , while in parallel field the opposite occurs. The orientations of bias current, applied field and GB (dashed lines) in perpendicular and parallel field configurations are also illustrated.

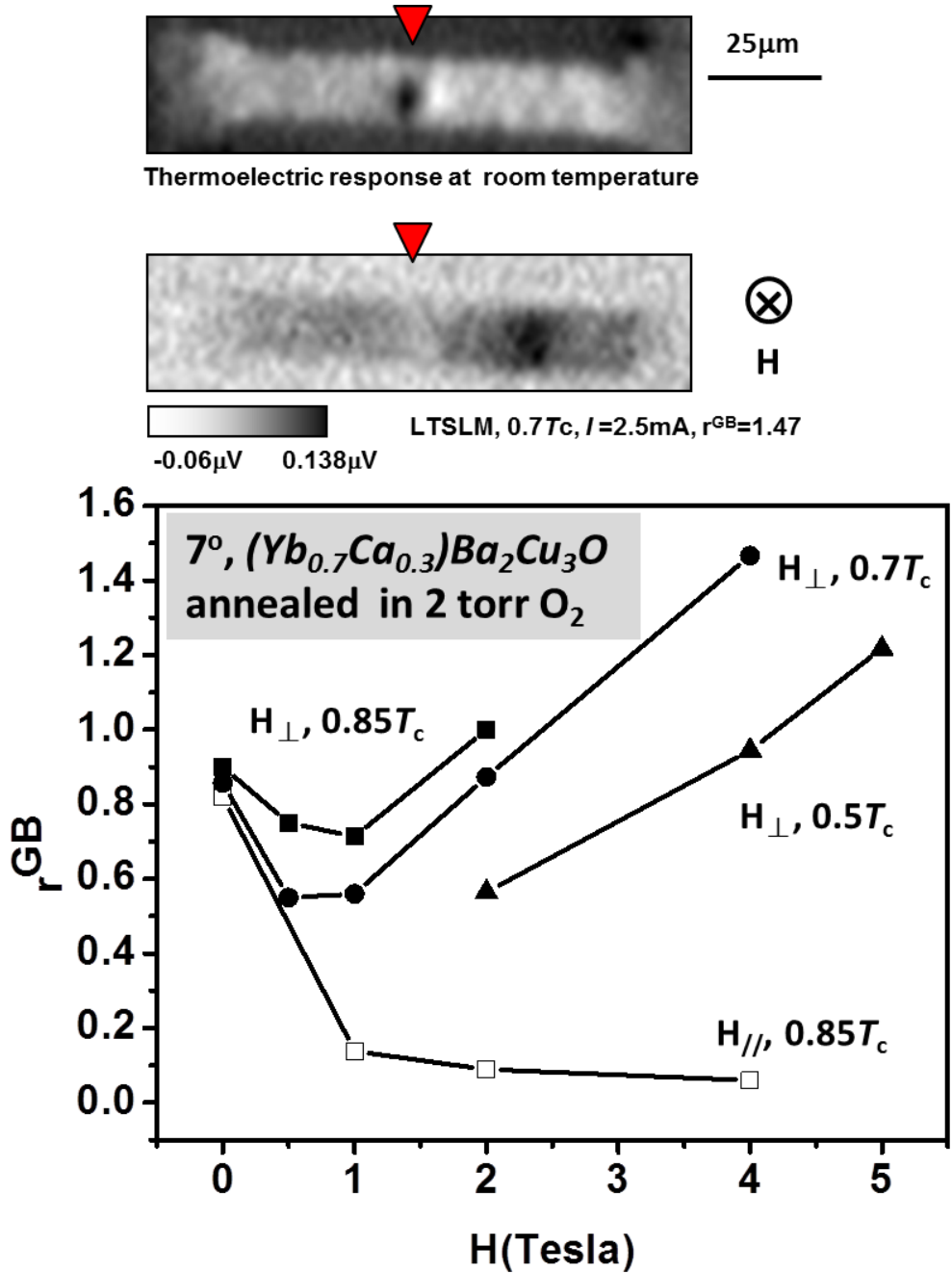

Figure 9. An oxygen-deficient $(Yb_{0.7}Ca_{0.3})BCO$ 7° bi-crystal sample R-0.3. The sample was first annealed in 1 bar oxygen to achieve a stable $T_c$ and then annealed for 24 hours in 2 torr oxygen. The $T_c$ of the sample was 71.2K and $J_c^{grain}$ (0.85$T_c$, self-field) was 0.54MA/cm². Top: the LTSLM image at 4T (perpendicular field), 0.7 $T_c$ shows a transparent GB because the $\delta V$ response is generated everywhere in the track except at the GB. The position of the GB is marked with red arrows. Bottom: $r^{GB}$ =$J_c^{GB}/J_c^{grain}$ plotted as a function of reduced temperature $t$ and applied field. In perpendicular field, $r^{GB}$ increases with increasing field. In parallel field, $r^{GB}$ decreases rapidly with increasing field. The definition of perpendicular and parallel field configuration is the same as in Figure 8.

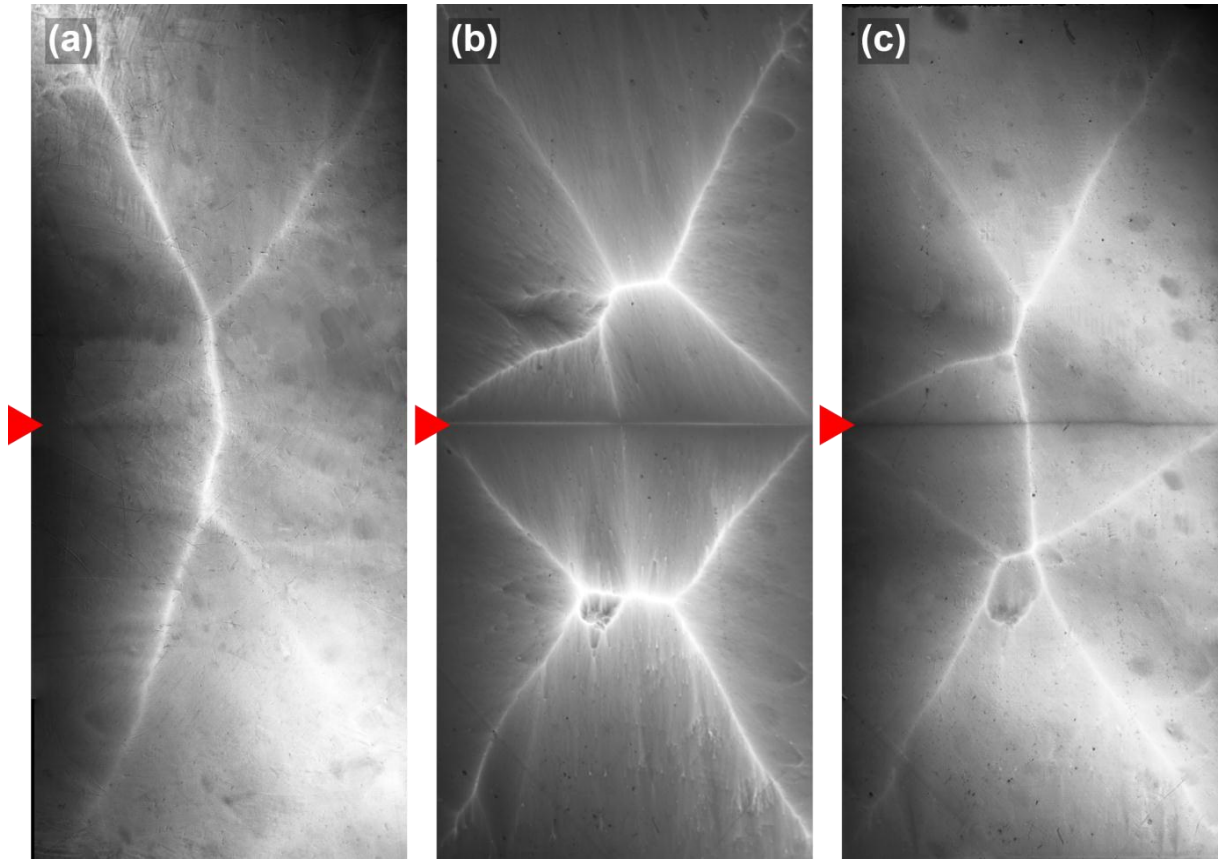

Figure 10. Field-cooled (FC) MOI images taken at 10K of a same piece of 6º, 30% Ca-doped YbBCO bi-crystal films at different stages of annealing at 475ºC: (a) after 2 hours annealing in 1 bar oxygen, $J_c^{GB} \approx J_c^{grain}$=2.8MA/cm$^2$; (b) after 36 hours annealing in 1 bar oxygen, $J_c^{GB}$=4.6MA/cm$^2$, $J_c^{grain}$=10MA/cm$^2$; (c) after 24 hours annealing in 500mtorr oxygen, $J_c^{GB}$=1.9MA/cm$^2$, $J_c^{grain}$=2.8MA/cm$^2$. The position of the GB is marked with the red arrows. $J_c^{grain}$ and $J_c^{GB}$ are calculated by quantitative analysis of MOI images [17]. While extensive 1 bar oxygen annealing increased both $J_c^{GB}$ and $J_c^{grain}$, it also reduced r$^{GB}$. On the contrary, annealing under a reduced oxygen partial pressure depressed both $J_c^{GB}$ and $J_c^{grain}$ but improved r$^{GB}$. Refer to text for more information.

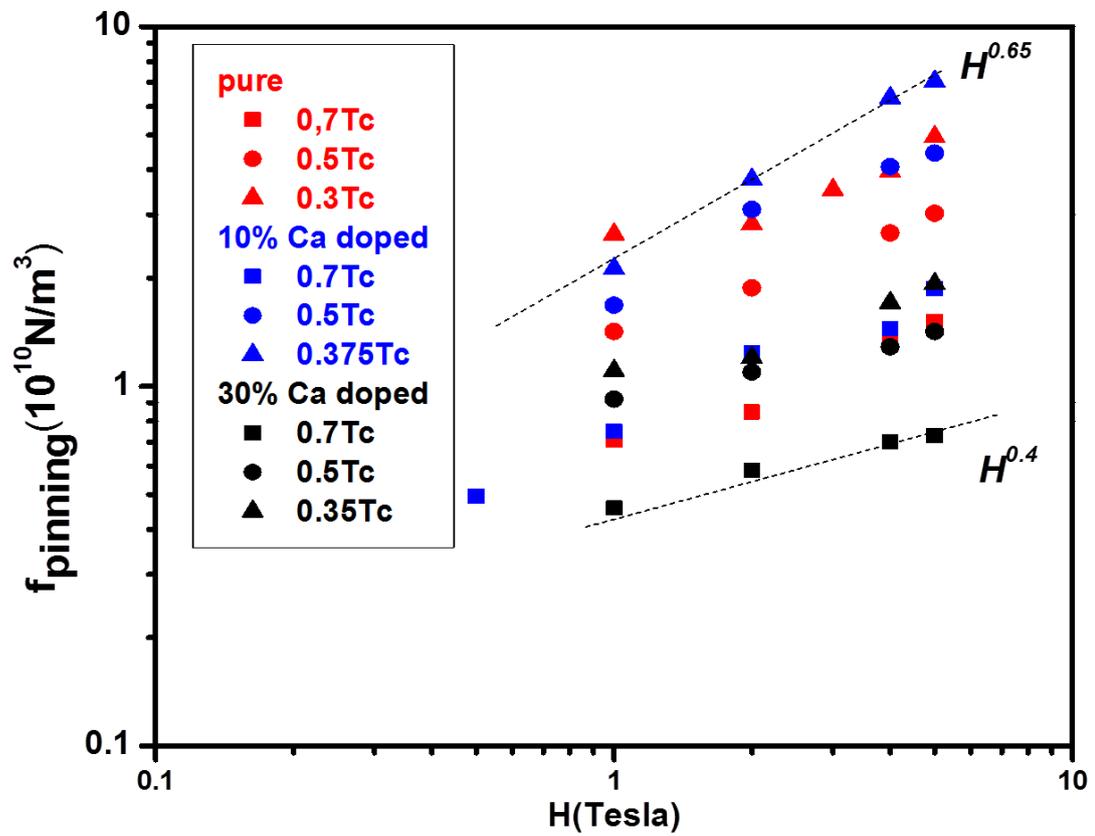

Figure 11. Pinning force density of 7° GBs in S-0, S-0.1 and S-0.3. The fitted data show that pinning force density $F_{pinning}$ scales as $H^\alpha$ with $0.4<\alpha<0.65$, which is consistent with pinning by GB dislocations[33, 34].

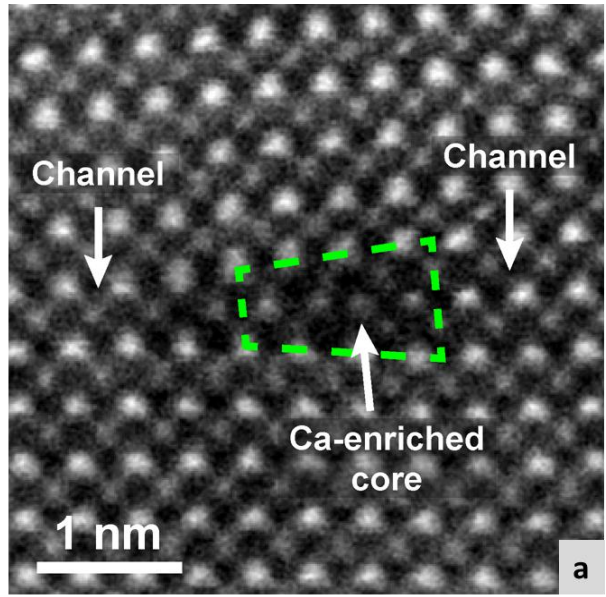

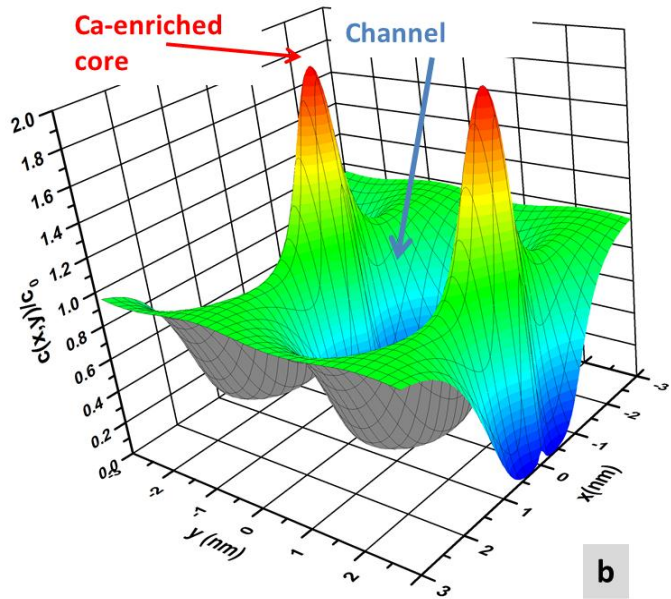

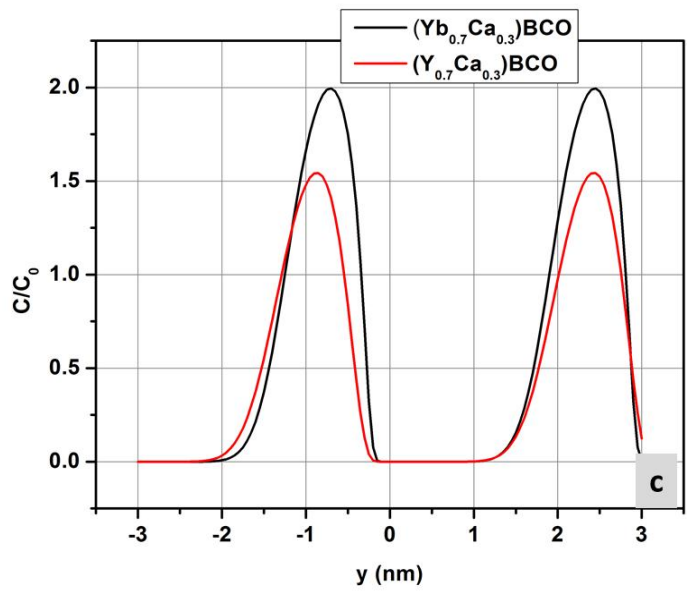

Figure 12. (a) High angle annular dark field scanning transmission electron microscope (HAADF-STEM) image showing the atomic structure of GB dislocation in a $(Yb_{0.7}Ca_{0.3})$BCO bi-crystal. The green dashed line indicates the dislocation core region. (b) Calculated Ca distribution $c(x,y)$ near a 7º GB in $(Yb_{0.7}Ca_{0.3})$BCO based on the segregation model[7]. $c(x,y)$ is renormalized to bulk Ca-doping level $c_0$, which is 0.3 here. Driven by local strain field and charge imbalance near GB, Ca segregates to the core regions while the channels between cores are low in Ca [7, 11-15]. (c): Ca concentration along the GB plane ($x=0$), $c(0,y)$, of a 7º [001] tilt GB in $(Yb_{0.7}Ca_{0.3})$ and $(Y_{0.7}Ca_{0.3})$ films. Due to the greater volume difference between $Yb^{3+}$ and $Ca^{2+}$, the peak Ca concentraion in (YbCa)BCO is higher than that in (YCa)BCO. On the other hand, the width of Ca-poor channels is almost the same. The fitting parameters used to generate this plot are the same as those used in Ref[7].

Table 1. Optimized PLD growth condition for $(Yb_{1-x}Ca_x)BCO$ thin films. The typical substrate to target distance was 55mm.

| Substrate | $x$ | Heater temperature (°C) | Oxygen pressure (mtorr) |
|---|---|---|---|
| STO (single crystal and bi-crystal) | 0 | 800 | 450 |
| | 0.1 | 810 | 300 |
| | 0.3 | 810 | 200 |

Table 2. Details on the five 7° bi-crystal films studied by LTSLM. Extensive annealing means that further annealing under same condition did not cause any further $T_c$ change. All annealing was done at 475°C.

| sample | Ca doping level $x$ | Annealing history | $T_c$(K) | $J_c^{grain}$(MA/cm²), at $0.85T_c$ in self-field. |
|---|---|---|---|---|
| T-0.1 | 0.1 | 2 hours in 1 bar oxygen | 61.3 | 0.14 |
| S-0 | 0 | extensively in 1 bar oxygen | 85.0 | 1.48 |
| S-0.1 | 0.1 | extensively in 1 bar oxygen | 79.2 | 1.80 |
| S-0.3 | 0.3 | extensively in 1 bar oxygen | 75.8 | 1.21 |
| R-0.3 | 0.3 | extensively in 1 bar and then in 2 torr oxygen | 71.2 | 0.54 |